\documentclass[showpacs,reprint,aps, prl, longbibliography]{revtex4-1}%
\usepackage{amsmath}
\usepackage{graphicx}
\usepackage{amsfonts}
\usepackage{amssymb}
\usepackage{setspace}
\usepackage{color}
\usepackage{dcolumn}
\usepackage{bm}
\usepackage{multirow}

\begin{document}

\title{Breaking the cavity linewidth limit of resonant optical modulators}
\author{Wesley D. Sacher$^{1*}$}
\author{William M. J. Green$^{2*}$}
\author{Solomon Assefa$^{2}$}
\author{Tymon Barwicz$^{2}$}
\author{Huapu Pan$^{2}$}
\author{Steven M. Shank$^{3}$}
\author{Yurii A. Vlasov$^{2}$}
\author{Joyce K. S. Poon$^{1}$}

\thanks{\textbf{Corresponding authors:} J. K. S. Poon, W.  M. J. Green, and W. D. Sacher}
\thanks{\textbf{Contributions:} W.D.S. performed the measurements.  W.D.S., W.M.J.G., and J.K.S.P. conceived and designed experiments and analyzed the data. W.D.S. and J.K.S.P. performed the analytical modeling of the modulators.  W.M.J.G., S.A., T.B., S.M.S., and Y.A.V. developed the modulator integration  process and qualified the devices with the help of H.P..  W.D.S., W.M.J.G., S.A., Y.A.V. and J.K.S.P. wrote the paper.\\}

\affiliation{$^1$Department of Electrical and Computer Engineering and Institute for Optical Sciences, University of Toronto, 10 King's College Road, Toronto, Ontario, M5S 3G4, Canada}
\affiliation{$^2$IBM Thomas J. Watson Research Center, 1101 Kitchawan Road, Yorktown Heights, New York 10598, U.S.A.}
\affiliation{$^3$IBM Systems and Technology Group, Microelectronics Division, 1000 River St., Essex Junction, Vermont 05452, U.S.A. }

\date{\today }

\maketitle

\textbf{Microring optical modulators are being explored extensively for energy-efficient photonic  communication networks in future high-performance computing systems and microprocessors  \cite{CoteusIBM2011, CunninghamJSTQE2011}, because they can significantly reduce the power consumption of optical transmitters via the resonant circulation of light  \cite{XuNAT2005, ZortmanCLEO2010, DongOL2009}.   However, resonant modulators have traditionally suffered from a trade-off between their power consumption and maximum operation bit rate, which were thought to depend oppositely upon the cavity linewidth \cite{GheormaJLT2002, RabieiJLT2002, MillerIEEEProc2009}.  Here, we break this linewidth limitation using a silicon microring. By controlling the rate at which light enters and exits the microring, we demonstrate modulation free of the parasitic cavity linewidth limitations at up to 40 GHz, more than $6\times$ the cavity linewidth.  The device operated at 28 Gb/s using single-ended drive signals less than 1.5 V.  The results show that high-$Q$ resonant modulators can be designed to be simultaneously low-power and high-speed, features which are mutually incompatible in typical resonant modulators studied to date.}

\begin{figure}[b]
\centering
\includegraphics[width=89 mm]{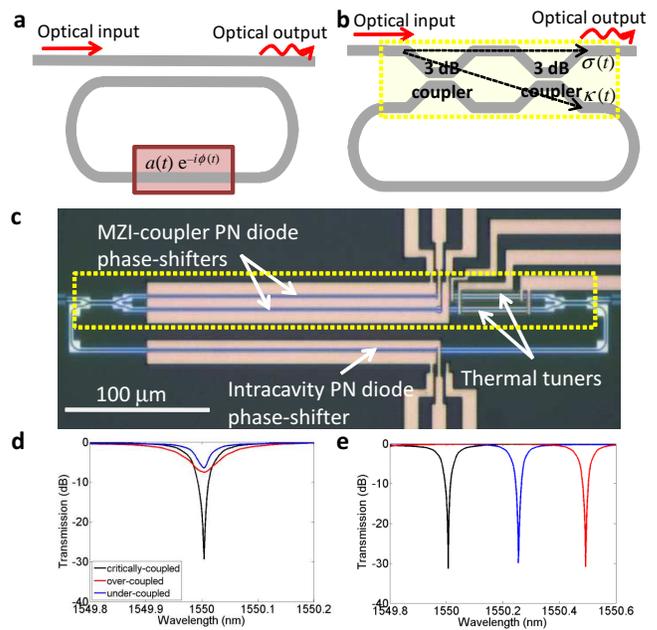}
\caption{Schematics, image, and tunable transmission spectra of the microring modulator. (a) Schematic of an intracavity modulated microring. (b) Schematic of a coupling modulated microring that uses a $2 \times 2$ MZI-coupler as marked by the box. (c) Optical microscope image of the fabricated SOI microring with the $2 \times 2$ MZI-coupler marked by the box.  Measured transmission spectra for (d) tuning the coupling coefficient at a fixed resonance and (e) tuning the resonance wavelength with a fixed coupling coefficient.   Independent tuning of the coupling and resonance wavelength using the thermal tuners was achieved.}\label{fig:microring}
\end{figure}

A microring optical modulator, in its simplest form, consists of a closed waveguide loop coupled to an input/output waveguide.  A constant light beam is the input, and a physical parameter of the microring is modulated to produce a time-varying optical output. Two distinct operation modes are intracavity and coupling modulation. The vast majority of microring modulators to date uses intracavity modulation (Fig. \ref{fig:microring}a), where the circulating optical field is modulated by the intracavity round-trip phase, $\phi(t)$, and/or loss, $a(t)$, while the coupler parameters are fixed  \cite{RabieiJLT2002, XuNAT2005, ZortmanCLEO2010, DongOL2009}.  Because the intracavity optical field amplitude rises and falls at a time-scale set by the photon cavity lifetime, the maximum intracavity modulation bandwidth diminishes with increasing $Q$ \cite{SacherOE2008}. Additionally, complete on/off modulation (0-100\% transmission)  requires the stored intracavity optical energy be completely charged and depleted in each switching cycle.  Thus, whether in the small- or large- output signal regime, the intracavity modulation bandwidth is inherently limited by the cavity linewidth.

Coupling modulation circumvents this linewidth limitation. In coupling modulation, the intracavity parameters, $\phi$ and $a$, are kept constant, while the through- and cross-coupling coefficients, $\sigma(t)$ and $\kappa(t)$ respectively, are modulated (Fig. \ref{fig:microring}b) \cite{SacherOE2008, SacherJLT2009}. We term the regime where the modulation rate is greater than the cavity linewidth as ``non-adiabatic coupling modulation'' \cite{SacherJLT2009}. ``Adiabatic coupling modulation''  refers to modulation rates less than the linewidth, when the intracavity optical field in its entirety follows the modulation signal \cite{YarivPTL2002, GreenOE2005, ZhouOE2007,GillJSTQE2010}.  Distinct from the adiabatic regime, intracavity modulation, and $Q$-switching \cite{DongOL2009b},  non-adiabatic coupling modulation does not completely dump the stored intracavity optical energy to generate near 0-100\% transmission swings. Instead, it extracts, in the transient, minor fractions of the large intracavity optical field in a high finesse cavity to produce output optical pulses with peak powers that can equal the input optical power. The coupler gates the intracavity optical field as it exits the microring to enable the bandwidth of non-adiabatic coupling modulation to exceed the cavity linewidth \cite{SacherJLT2009}. Non-adiabatic coupling modulation is a resonantly enhanced phenomenon, since the required changes to the coupling coefficients, and hence the device power consumption, reduce as the stored intracavity optical energy increases \cite{SacherJLT2009}.

In this work, we demonstrate and investigate non-adiabatic coupling modulation using a silicon microring incorporating a $2 \times 2$ Mach-Zehnder interferometer (MZI) as a coupler \cite{YarivPTL2002, GreenOE2005, ZhouOE2007, GillJSTQE2010}, as illustrated in Fig. \ref{fig:microring}(b). The MZI-coupler provides independent control of the coupling coefficient and resonance wavelength.  The differential phase-shift between the MZI-coupler arms changes $\kappa(t)$ and $\sigma(t)$ \cite{YarivPTL2002, GreenOE2005}, while the common-mode phase-shift changes the round-trip optical path length.  Figure \ref{fig:microring}(c) shows the microring fabricated on a silicon-on-insulator (SOI) substrate using the IBM  Silicon CMOS Integrated Nanophotonics process \cite{AssefaOFC2011}. The MZI had 3 dB directional couplers, 50 $\mathrm{\mu m}$ long thermal tuners, and matched  200 $\mathrm{\mu m}$ long PN diode phase-shifters for push-pull modulation.   An identical PN diode phase-shifter was included inside the microring to facilitate direct comparisons between intracavity and coupling modulation with the same device.  A  MZI with identical PN diode phase-shifters was fabricated as a reference to extract the electro-optic characteristics of the MZI-coupler independent of the microring (see Supplementary Information). The static transmission spectra of the microring in Fig. \ref{fig:microring}(d)-(e) demonstrate the precise and independent tuning of the coupling coefficient and resonance wavelength achieved by the thermal tuners.  An extinction ratio near 30 dB was reached at critical coupling.  With no voltage applied to the PN diode phase-shifters, the cavity linewidth at critical coupling was $\Delta \nu \approx 7$ GHz, corresponding to a loaded $Q$ of about 28000 and a finesse of $14$.

\begin{figure}[t]
\centering
\includegraphics[width=89 mm]{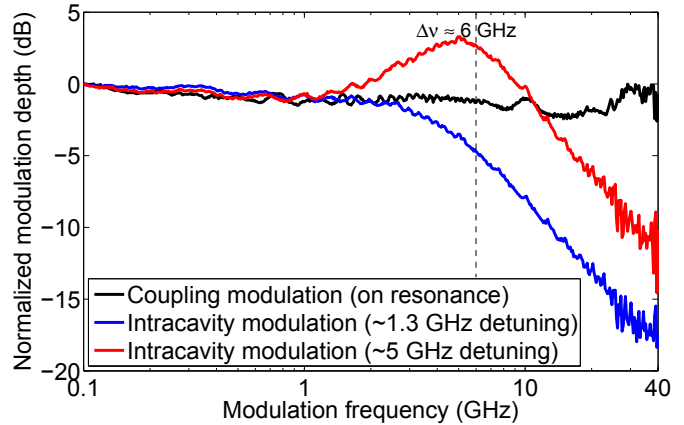}
\caption{Optical small-signal modulation responses of coupling and intracavity modulation.  Each curve is obtained by normalizing the electro-optic $S_{21}$ of the microring to the $S_{21}$ of the reference MZI and referencing to the value at 100 MHz.  The microring was biased near critical coupling, with a cavity linewidth $\Delta \nu \approx 6$ GHz. The intracavity modulation response for a $\sim 1.3$ GHz detuning from resonance (blue) has a 3 dB bandwidth of 4.4 GHz, similar to the linewidth.  A $\sim 5$ GHz detuning produces a resonant sideband peak near the value of the detuning (red), and the 3 dB bandwidth is extended to $\sim 13$ GHz.  The coupling modulation response (black) does not roll-off to 40 GHz (more than $6\times$ the linewidth). 
}\label{fig:smallsig}
\end{figure}

The small-signal optical modulation characteristics in Fig. \ref{fig:smallsig} demonstrate that the coupling modulation bandwidth significantly exceeds the traditional cavity linewidth limit. In the figure, the resonant optical modulation response was isolated from the electrical characteristics of the measurement setup and  PN diode phase-shifters by normalizing the electro-optic $S_{21}$ parameter of the microring modulator to the $S_{21}$ of the reference MZI (details in the Supplementary Information).  The phase-shifters were operated in depletion with a DC bias of -1 V.  Under bias, the device was near critical coupling and had a cavity linewidth of $\Delta \nu \approx 6$ GHz.  The input laser wavelength was on resonance for coupling modulation and was detuned slightly off-resonance for intracavity modulation to obtain an appreciable modulation depth. The black curve in Fig. \ref{fig:smallsig} shows that in contrast to intracavity modulation, the coupling modulation response did not roll off to 40 GHz, more than $6\times$ the cavity linewidth. The maximum frequency measured here was limited by the instrumentation.  The flat response indicates that the frequency roll-off characteristics of the non-adiabatic coupling modulation resembled those of the non-resonant reference MZI. The slight decrease in the modulation depth near the frequency corresponding to the cavity linewidth was due to a slight under-coupling of the microring \cite{SacherOE2008}.  

In comparison, the intracavity modulation response with input light that is  1.3 GHz detuned  from resonance (blue curve) had a 3 dB bandwidth of about 4.4 GHz, confirming that the intrinsic  modulation bandwidth was limited by the cavity linewidth. For larger detunings, a peak occurs when a modulation sideband is on resonance and becomes comparable to or greater than the carrier in amplitude within the microring \cite{SacherOE2008,GillJSTQE2010}. The peak is exaggerated for large detunings, because the circulating amplitude of an off-resonant carrier is small.  This effect is shown by the $\sim 5$ GHz detuning measurement (red curve).  Although the modulation sideband peak extended the 3 dB bandwidth to about 13 GHz, intracavity modulation of a highly detuned carrier is not practical, because the absolute modulation depth and the linearity of the modulator are compromised.

\begin{figure}[!ht]
\centering
\includegraphics[width=87 mm]{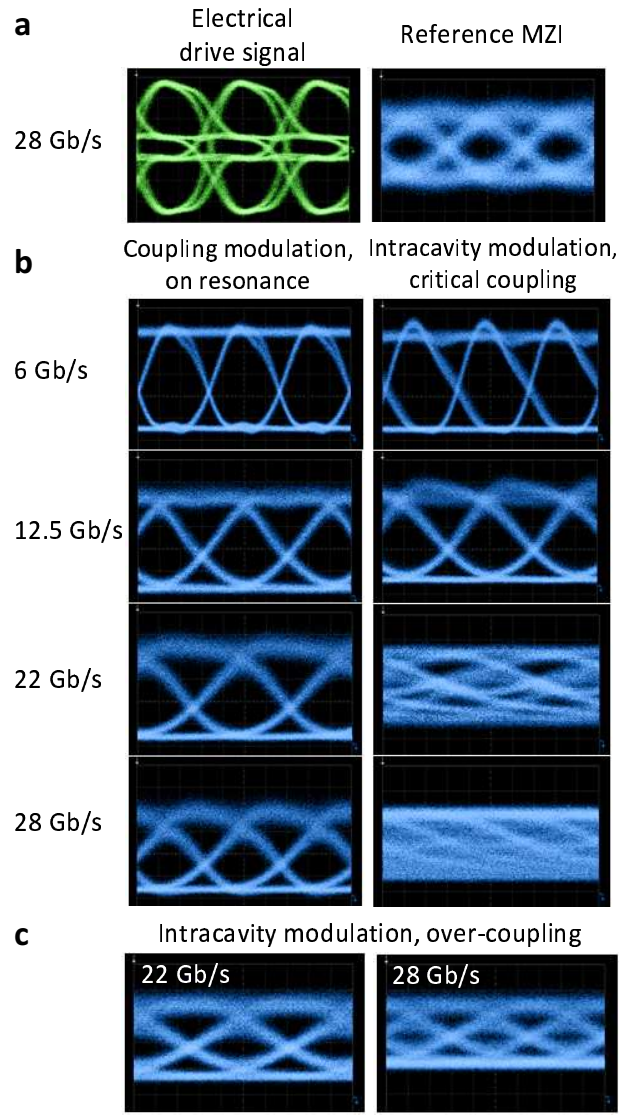}
\caption{Eye diagrams of large-signal coupling and intracavity modulation. (a) Eye diagrams of the pre-emphasized electrical drive signals at 28 Gb/s (left) and the resultant optical output of the reference MZI (right). No remnants of the pre-emphasis are present in the optical output. (b) Coupling (left) and intracavity (right) modulation eye diagrams at 6-28 Gb/s for bias points near critical coupling ($\Delta \nu \approx 6-7$ GHz).  The coupling modulation eye is open at 28 Gb/s, but the intracavity modulation eye closes at bit rates greater than roughly $2\times$ the linewidth. (c) Intracavity modulation eye diagrams of an over-coupled microring ($\Delta \nu \approx 9$ GHz).  The eye opening is larger than in (b), confirming that the intracavity modulation bandwidth depends on the cavity linewidth.}\label{fig:eye}
\end{figure}

Large-signal data modulation and optical eye diagram measurements provide further evidence that the coupling modulation bandwidth is not similarly limited by the cavity linewidth as intracavity modulation. To reduce the required drive voltage, the PN diode phase-shifters were in forward bias, with a DC offset of 0.28 V, and driven by pre-emphasized non-return-to-zero (NRZ) $2^{31}-1$ pseudo-random bit sequence (PRBS)  signals. The electrical pre-emphasis extended the modulation bandwidth of the PN diode phase-shifters far beyond their minority carrier lifetime limit of $\sim 1$ GHz \cite{XuOE2007,GreenOE2007,ManipatruniLEOS2007}.  The single-ended voltage swing of the pre-emphasized bits was 1.5 $\mathrm{V}_{pp}$, and the non-emphasized bits were between 0.24-0.3 $\mathrm{V}_{pp}$.  Fig. \ref{fig:eye}(a) (left) shows the eye diagram of the pre-emphasized 28 Gb/s drive signal. The optical output of the reference MZI driven with this signal in push-pull mode  (Fig. \ref{fig:eye}(a), right) shows that the MZI-coupler optical output did not contain remnants of the pre-emphasis which could potentially extend the microring modulation bandwidth beyond the limits of the resonant optical dynamics.

Figure \ref{fig:eye}(b) summarizes the coupling and intracavity modulation eye diagrams of the microring at bit rates between 6 and 28 Gb/s. The cavity linewidths at the operating bias points were $\Delta \nu \approx 6-7$ GHz. At each bit rate, identical drive signals were applied to the coupler or intracavity phase-shifters, except the MZI-coupler was driven in push-pull while the intracavity phase-shifter was driven single-ended.   For coupling modulation, the microring was modulated between critical coupling and under-coupling with the input light on resonance.  For intracavity modulation, the microring was biased at critical coupling and the input wavelength was slightly detuned from resonance.  The pre-emphasis ratio and detuning were optimized to maximize the eye opening for each case. At 6 and 12.5 Gb/s, both the coupling and intracavity modulation eye diagrams had extinction ratios of 10-13 dB and a maximum optical transmission $> 40\%$.  As the bit rate increased, the coupling modulation eye remained wide open up to 28 Gb/s, whereas the intracavity modulation eye was closed at 22 Gb/s.  However, because of the modest finesse and PN diode phase-shifter efficiency, the extinction ratio of coupling modulation at 22 and 28 Gb/s decreased to 10 dB, and the maximum optical transmission was only about 10 to 20\% of the off-resonance transmission.  

To check that the intracavity modulation eye closure was due to the cavity linewidth and not to an electrical artifact, we over-coupled the microring to increase its linewidth to 9 GHz at the expense of modulator efficiency and extinction ratio. Fig. \ref{fig:eye}(c) shows that the eye opening increased at 22 Gb/s, but remained closed at 28 Gb/s. The eye diagrams show that intracavity modulation suffered from severe inter-symbol interference (ISI) at bit rates greater than roughly $2\times$ the cavity linewidth (i.e. $> 12.5$ Gb/s), while coupling modulation at up to $4\times$ the linewidth was not similarly affected. Coupling modulation should function well beyond 28 Gb/s; however, the measurements were limited by the instrumentation.  The results in Fig. \ref{fig:smallsig} and \ref{fig:eye} demonstrate that the long-held cavity linewidth limit to the intracavity modulation bandwidth can be broken with coupling modulation.

\begin{figure}[!ht]
\centering
\includegraphics[width=89 mm]{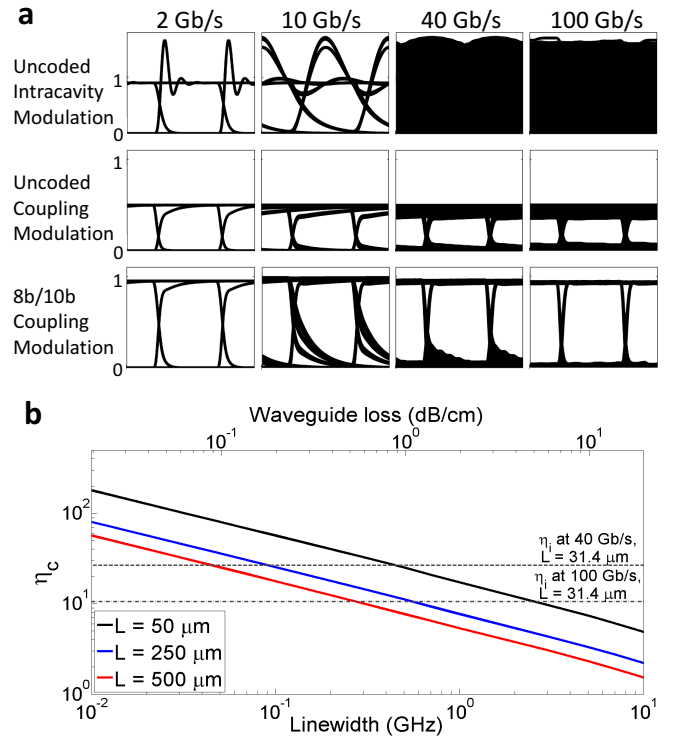}
\caption{ISI and efficiency scaling of coupling and intracavity modulation.  The calculations assume a group index of 4.3, a NRZ PRBS $2^{17}-1$ data signal, a resonant input for coupling modulation, and critical coupling for intracavity modulation.  (a) Computed eye diagrams at several bit rates assuming $\Delta\nu = 5$ GHz and a round-trip length of 250 $\mathrm{\mu m}$ for  (top) intracavity modulation and (center) coupling modulation driven by an uncoded NRZ signal, and (bottom) coupling modulation driven by a NRZ 8b/10b encoded signal. With DC-balanced encoding, coupling modulation can achieve a 0-90\% swing at 100 Gb/s.  (b) The coupling modulation efficiency, $\eta_c$, versus microring waveguide loss and cavity linewidth for several round-trip lengths, $L$. The calculations assume a 8b/10b encoded drive signal and a 0-90\% output swing.  The intracavity efficiency, $\eta_i$, of a 5 $\mathrm{\mu m}$ radius microring with the same output swing at 40 Gb/s and 100 Gb/s, using linewidths of 20 GHz and 50 GHz respectively, are marked for comparison.  Coupling modulation becomes increasingly efficient over intracavity modulation as the $Q$ factor and bit rate increase.  }\label{fig:scaling}  
\end{figure}

A potential drawback to coupling modulation is the ISI originating from the \emph{low} frequency content of the drive signal, which can significantly deplete the stored optical energy in the cavity.  Fortuitously, this ISI can be mitigated.  One suggestion is to modulate two output couplers to maintain a constant intracavity optical power  \cite{PopovicIPR2010} at the expense of device complexity, cavity finesse, and power efficiency for large-signal modulation.  Here, we propose a more direct approach of encoding the electrical data to produce a DC-balanced drive signal.  An example is the 8b/10b code, a typical line code for the Ethernet and InfiniBand communication standards.  To illustrate the effect of the ISI and encoding,  Fig. \ref{fig:scaling}(a) compares a series of computed eye diagrams.  The simulations use a NRZ PRBS $2^{17}-1$ pattern, a 5 GHz cavity linewidth, and a 250 $\mathrm{\mu m}$ round-trip length.  The top row shows that uncoded intracavity modulation requires a linewidth that is at least half the bit rate, in agreement with our experimental results.  The middle row shows that for uncoded coupling modulation in the non-adiabatic regime, the transmission swing in the eye opening can be about 25\%. Larger drive signals would increase the ISI.  As in our experiment, the resonator is driven between under-coupling and critical coupling. The bottom row shows that with an 8b/10b encoded drive signal, non-adiabatic coupling modulation with an eye opening of about 90\%  and low ISI are possible at 100 Gb/s, characteristics that cannot be attained by intracavity modulation.  Significantly, the ISI in coupling modulation diminishes as the bit rate increases, since the low frequency content of the modulation signal is reduced.

Although our results demonstrate that the coupling modulation bandwidth can be substantially larger than the intracavity modulation bandwidth, an essential question remains of whether coupling modulation of a narrow linewidth resonator can be more power efficient than  intracavity modulation of a small resonator with a broad linewidth.  Here, we find the general efficiency scaling relationships to show that coupling modulation can indeed be more efficient, particularly for high bit rate and large-signal modulation. Thus, coupling modulation can be suitable for energy-efficient large capacity optical links. 

We define an efficiency metric, $\eta = \Delta \phi_{MZI}/\Delta \phi_{ring}$, where $\Delta \phi_{ring}$  and $\Delta \phi_{MZI}$ are respectively the phase-shifts of a microring and a MZI biased at quadrature required to produce the same output transmission swing, assuming identical phase-shifters.  $\Delta \phi$ is single-ended for intracavity modulation and is applied push-pull as $\pm \Delta \phi/2$ for coupling modulation.  Referencing to $\Delta \phi_{MZI}$ allows for a general comparison between coupling and intracavity modulation independent of material platforms.  The phase-shifts are related to the power consumption of the device by the physical mechanism of electro-optic modulation and any drive circuitry.

For intracavity modulation, from the microring transmission function \cite{YarivPTL2002}, we find the efficiency, $\eta_i$, is roughly proportional to the \emph{intracavity power} or finesse, $F$: 
\begin{equation}\label{eqn:etai}
\eta_i \approx k_i F,
\end{equation}
where $k_i \lesssim 0.42$ depends on the high and low transmission levels as well as the ratio of the round-trip loss to the coupling.  As examples, at critical coupling and $F \gtrsim 5$, $k_i = 0.24$ for a 0-90\% output swing and $k_i = 0.41$ for a 20-30\% swing.  $k_i$ is  lower for large-signal modulation than in the small-signal regime because the microring transmission spectrum flattens at wavelengths detuned from the resonance.  In contrast, the efficiency of coupling modulation, $\eta_c$, scales with the \emph{intracavity field}. For a MZI-coupler and a resonant optical input, 
\begin{equation}\label{eqn:etac}
\eta_c \approx k_c\sqrt{F},
\end{equation}
where $k_c \lesssim 1$  depends on the transmission levels and  $F$ is the finesse at critical coupling.  For the same transmission levels, $k_c$ is lower in the non-adiabatic than adiabatic regime due to a reduced intracavity optical field amplitude.  In the adiabatic regime and for $F \gtrsim 5$, we find $k_c = 0.85$ for a 0-90\% output swing and $k_c = 0.73$ for a 20-30\% swing.  In the non-adiabatic regime and $F \gtrsim 20$, $k_c = 0.41$ for a 0-90\% output swing.  It  follows from Eq. \ref{eqn:etai} and \ref{eqn:etac} that adiabatic coupling modulation is more efficient than intracavity modulation in cavities with the same $F$ and phase-shifters when $F \lesssim (k_c/k_i)^2$. While intracavity modulation is more efficient in the small-signal regime if the resonator has at least a moderate finesse (e.g. 20-30\% swing, $F \gtrsim 5$), adiabatic coupling modulation is more efficient for large-signal modulation as $k_i$ diminishes.  For example, $\eta_c > \eta_i$ for a 0-90\% output swing when $F \lesssim 10$ and for a 0-99\% swing when $F \lesssim 87$.

The efficiency scaling becomes especially favourable to non-adiabatic coupling modulation over intracavity modulation at high $Q$ factors and high bit rates.  Because the minimum cavity linewidth is dictated by the desired modulation rate, improvements in $\eta_i$ of intracavity modulation via the finesse must come from the cavity size reduction. However, a small cavity size prohibits the inclusion of tuning structures needed to attain large extinction ratios and large-signal swings in practice.  Thus, the key advantage of coupling modulated microrings is that they can be kept larger (to accommodate tunable couplers), while $\eta_c$ can, in principle, be arbitrarily boosted by increasing $Q$ to raise the finesse.  Figure \ref{fig:scaling}(b) shows  $\eta_c$ versus the cavity linewidth for several round-trip lengths assuming a 0-90\% output swing.  The values of $\eta_i$ for  SOI microrings with a 5 $\mathrm{\mu m}$ radius and the same output swings at 40 Gb/s and 100 Gb/s are marked.  As the round-trip length of a coupling-modulated microring increases from 50 $\mathrm{\mu m}$ to 500 $\mathrm{\mu m}$, a narrower linewidth is required for $\eta_c > \eta_i$.  

By combining the benefits of resonant enhancement with the large bandwidths of non-resonant modulators, coupling modulation, for the first time, opens the avenue toward ultra-low power yet high-speed modulation of ultra-high-$Q$ resonators.  Such resonators on silicon chips can possess finesses exceeding 10000  \cite{LeeNP2012}. 

\section{Methods}

The optical modulators were fabricated using the IBM Silicon CMOS Integrated Nanophotonics process \cite{AssefaOFC2011} on a 200 mm-diameter SOI wafer with a 2 $\mathrm{\mu m}$-thick buried-oxide layer and a 220 nm-thick top silicon layer. Fully-etched silicon access waveguides and partially-etched PN diode rib waveguides were defined and planarized with silicon dioxide through a shallow trench isolation module. Typical CMOS ion implantation conditions were used to form a lateral PN diode junction at the center of each phase-shifter. The diode junction was designed with a nominal carrier concentration of $5\times 10^{17}$ cm$^{-3}$ in the P- and N-type regions. After a rapid thermal activation anneal, silicide ohmic contacts to the phase-shifters were formed. The silicide also formed the resistive thermal tuners for the MZI bias control. Finally, tungsten vias and copper metal interconnects were formed to electrically contact the phase-shifters and thermal heaters. Dies were prepared with cleaved facets for on-/off-chip optical coupling using tapered optical fibers. 

The measurements were taken with a swept wavelength tunable laser (Agilent 81682A).  A 67 GHz vector network analyzer (VNA, Agilent E8361A) was used for the  measurements in Fig. \ref{fig:smallsig}.  For intracavity modulation, a single-ended signal was applied to the intracavity phase-shifter.  For coupling modulation, the MZI phase-shifters were driven in push-pull. To generate a differential drive signal, the VNA output was fed into a fanout circuit (Hittite HMC842LC4B), which had a bandwidth of about 32 GHz.  Custom 40 GHz RF probes (GGB Industries) contacted the devices. The output was detected with a 40 GHz InGaAs photoreceiver (Archcom Technology).

To generate the drive signals for the data modulation experiments of Fig. \ref{fig:eye}, the output of a PRBS generator (Centellax PPG12500, TG1P4A) was fed into a pre-emphasis converter (Anritsu MP1825B-002), which operated up to 28 Gb/s.  The pre-emphasized signals were directly fed to the RF probes.  The optical output of the modulator was amplified using an erbium doped fiber amplifier, bandpass filtered (full width at half maximum bandwidth of 0.8 nm), and captured on a digital communications analyzer with a 28 Gb/s optical module (Agilent 86100C, 86106B).  All eye patterns were obtained using $2^{31}-1$ PRBS and single-tap pre-emphasis.  

Analytic forms for $k_i$ and $k_c$ in Eq. \ref{eqn:etai} and \ref{eqn:etac} in the adiabatic regime were derived from the static transmission  \cite{YarivPTL2002}.  An analytic form for $k_c$ in the non-adiabatic regime was derived by assuming a periodic square-wave drive signal and solving for the average intracavity field with a rate equation.  This approximation neglects the exact shape  and low-frequency content of the drive signal.   Figure \ref{fig:scaling} was calculated using the  microring modulator model  in \cite{SacherOE2008, SacherJLT2009}.

\section{Acknowledgments}

W.D.S. and J.K.S.P. thank the University of Toronto Emerging Communications Technology Institute for access to the VNA and RF signal generators. W.M.J.G. and W.D.S. thank D. M. Gill at IBM Research for helpful discussions. W.D.S. and J.K.S.P. are grateful for the financial support of the Natural Sciences and Engineering Research Council of Canada.

\bibliography{references}

\newpage
\renewcommand{\thefigure}{S\arabic{figure}}
\renewcommand{\theequation}{S\arabic{equation}}

\section{Supplementary information}

This  section provides details about the reference Mach-Zehnder interferometer (MZI)  and the small-signal modulation response ($S_{21}$) measurements.  The $S_{21}$ normalization procedure used in Fig. 2 is also discussed.

\subsection{Reference MZI}

\begin{figure}[b]
\centering
\includegraphics[width=87 mm]{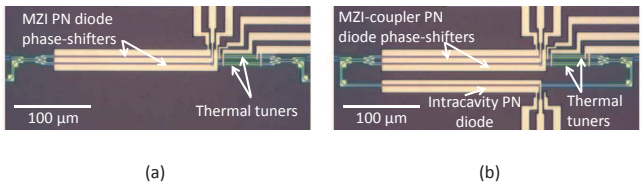}
\caption{Optical microscope images of (a) the reference MZI and (b) the microring modulator.}\label{fig:microscopePics}
\end{figure}

To isolate the modulation responses of intracavity and coupling modulation from the electrical characteristics of the PN diode phase-shifters and experimental setup, a reference MZI device was fabricated and measured.  Specifically, measurements of the reference MZI were used to normalize the $S_{21}$ measurements in Fig. 2 and to ensure the pre-emphasis in the pseudo-random bit sequence (PRBS) drive signals only equalized the roll-off due to the modulation bandwidth limitations of the PN diode phase-shifters and measurement setup in Fig. 3(b).  The reference MZI and microring modulator measured in this work are shown in Fig. \ref{fig:microscopePics}.  The reference MZI is nominally identical to the output coupler of the microring (i.e., designed to have identical waveguides, PN diode phase-shifters, thermal tuners, and wiring).  The reference MZI and microring modulator were on the same die and separated by about 620 $\mathrm{\mu}$m.

\subsection{$S_{21}$ measurements and normalization}

\begin{figure}
\centering
\includegraphics[width=87 mm]{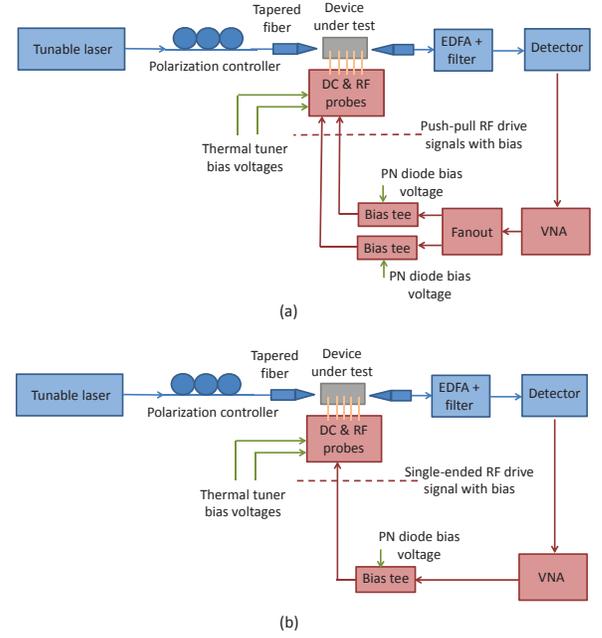}
\caption{Schematics of the $S_{21}$ measurement setups for (a) coupling modulation and push-pull modulation of the reference MZI and (b) intracavity modulation and single-arm modulation of the reference MZI.}\label{fig:setup}
\end{figure}

Push-pull reference MZI and coupling modulation electro-optical $S_{21}$ measurements were performed using the experimental setup illustrated in Fig. \ref{fig:setup}(a).  The $S_{21}$ parameter was collected from the voltage of the photodetector referenced to the vector network analyzer (VNA) output.  A high-speed fanout circuit split the single-ended RF output of the VNA into push-pull drive signals for the microring coupler or reference MZI.  Single-arm drive reference MZI and intracavity modulation $S_{21}$ measurements were performed using the experimental setup shown in Fig. \ref{fig:setup}(b).  Push-pull drive signals were not required for single-arm drive of the reference MZI and intracavity modulation, so the fanout was not included.  

The reference MZI $S_{21}$ measurements were performed at the quadrature bias point, which was established by applying DC voltages to the thermal tuners in the MZI arms.  Experimentally, it was found that the shape of the reference MZI $S_{21}$ curves was insensitive to changes in the MZI bias point.  The quadrature bias point was chosen to maximize the modulation efficiency of the MZI, and thus, reduce the relative noise in the $S_{21}$ measurements of the reference MZI.  The intracavity modulation $S_{21}$ measurements were taken at critical coupling with roughly 1.3 GHz and 5 GHz bias detunings.  The coupling modulation $S_{21}$ measurements were taken with the input wavelength on resonance and a slightly under-coupled bias.  In all cases, the PN diode phase-shifters were biased at $-1$ V.

\begin{figure}
\centering
\includegraphics[width=87 mm]{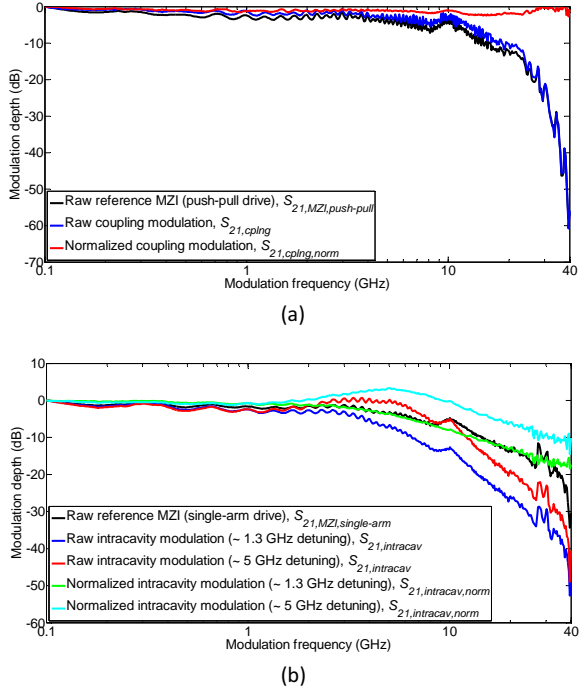}
\caption{$S_{21}$ measurements for (a) coupling modulation and push-pull modulation of the reference MZI and (b) intracavity modulation and single-arm modulation of the reference MZI.  $S_{21}$ measurements with only the RF cables, RF adapters, and bias tees de-embedded are shown for the microring and reference MZI (i.e., the raw $S_{21,cplng}$, $S_{21,MZI,push-pull}$, $S_{21,intracav}$, and $S_{21,MZI,single-arm}$).  Microring $S_{21}$ measurements normalized by the reference MZI response using  Eq. \ref{eqn:norm} are also shown (i.e., $S_{21,cplng,norm}$ and $S_{21,intracav,norm}$).}\label{fig:S21}
\end{figure}

The responses of the RF cables, RF adapters, and bias tees were de-embedded from the $S_{21}$ data; however, the responses of the fanout, RF probes, and on-chip wiring remained embedded in the $S_{21}$ data.  This raw $S_{21}$ data is shown in Fig. \ref{fig:S21} (i.e., the first two curves in (a) and the first three curves in (b)); each raw $S_{21}$ curve is referenced to its value at a 100 MHz modulation frequency.  $S_{21,MZI,push-pull}$ and $S_{21,MZI,single-arm}$ are defined as the raw push-pull and single-arm drive reference MZI responses, respectively.  $S_{21,cplng}$ and $S_{21,intracav}$ are defined as the raw responses of the coupling and intracavity modulation, respectively. To isolate the modulation responses of the optical cavity dynamics from those of the PN diode phase-shifters, on-chip wiring, RF probes, and fanout, the following normalization procedure was applied:
\begin{subequations}\label{eqn:norm}
\begin{equation}
S_{21,cplng,norm} = \frac{S_{21,cplng}}{S_{21,MZI,push-pull}},
\end{equation}
\begin{equation}
S_{21,intracav,norm} = \frac{S_{21,intracav}}{S_{21,MZI,single-arm}}.
\end{equation}
\end{subequations}   
$S_{21,cplng,norm}$ and $S_{21,intracav,norm}$ are the normalized $S_{21}$ data for coupling and intracavity modulation, respectively, and are presented in Fig. 2 and repeated in Fig. \ref{fig:S21}.  $S_{21,cplng,norm}$ and $S_{21,intracav,norm}$ represent the modulation responses of the microring due solely to optical cavity dynamics from coupling and intracavity modulation, respectively.

\end{document}